Photonic Framework to Handle Physical and Chemical Processes:
Quantum Entanglement, Coherence, De-coherence, Re-coherence and the
Roles of Multipartite Base States

O. Tapia

## Abstract

A quantum framework, according quantum theories of electromagnetic (EM) radiation to matter response, leads to a handy scheme addressed to examine both physical and chemical processes. Fundamental quantum effects such as entanglement, coherence, de-coherence and re-coherence yield a fully quantum physical presentation applicable to chemical processes.

A photonic basis-set obtains including all possibilities accessible to a system in abstract space; electronuclear (EN) base states put in resonance by photon fields are the ground where quantum states, q-states, would show time evolution. At laboratory space where energy and angular momentum conservation hold quantum dynamics (i.e. amplitude changes) is to be driven by low frequency EM radiation, e.g. microwaves. Materiality sustaining q-states unchanged.

Here, focus on construction and reading of photonic quantum states that underlie a host of physical processes recently submitted to experimental examination. The approach moves away interpretational issues using classical physics language to focus on apprehending quantum states including experimental information.

Department of Chemistry-Ångström
Uppsala University, Box 259, 75120 Uppsala, Sweden
Department of Physical Chemistry, Valencia University, Spain

email: orlando.tapia@kemi.uu.se

# 1. Introduction

The paper presents a quantum mathematical-chemistry framework including photon fields "dressing" elementary matter (molecular) states. The concept of photonic bases [1-3] ensures formulation of physical and chemical processes in full quantum physical terms; it starts up from abstract Hilbert spaces, thereafter incorporates photon fields states entangled to electronuclear (EN) bases sets to finally construct mappings to laboratory space rounding up an integrated quantum physical scheme where, and this is a novel feature, low frequency radiation calls to set up time evolution in coherent q-spaces.

Quantized electromagnetic (q-EM) states enter from the very beginning. Entanglements, coherence, decoherence and re-coherence in multi-partite abstract bases play central roles thereby opening a way to construct quantum physical descriptions [1-3] including chemical, photochemical as well as photo-biologic processes considered in their photonic wholeness.

In principle, the quantum framework is settled once all *response possibilities* towards external probing are taken into account: probing and probed instances are irreducible. The latter refers to a system under study the former is adapted to experimenter decisions. In abstract space, quantum physics is about possibilities; this grounding hypothesis underlies the abstract photonic scheme herein re-examined. [1-3] Probabilities may take over in laboratory space (Fence [4-6]), and only there, for reasons given later on.

The Fence, in our language, corresponds to a zone/place where one brings up abstract states and dress them with photon field so that one can physically probe, prepare, modulate or change the resultant laboratory quantum states, generically noted as q-states. [4] Materiality sustaining such q-states remains unchanged though being able to expressing possibilities infinite in number. The quantum states do not necessarily describe whereabouts of material elements, there is no representation involved; this is a clear difference with respect to classical physics style: q-states aren't objects.

Probing of such systems at laboratory level necessarily enters semi-classically via discrete events eliciting quantized energy exchanges; thus, in this scheme effective ("real") energy exchange materializes in probing (or laboratory measurements); this makes the difference compared to abstract level where operators may formally change abstract quantum states without any real energy expenses.

The dichotomy imposed by laboratory probing (measurement) permits casting q-states sustained by both material and pure electromagnetic components in a fully quantized manner. [4-6] In so doing, classical aspects of EM field fade away; it results in first constructing photonic basis with no interactions able to drive time evolution. This latter is empirically added via microwave radiation or any other low frequency radiation. This results in a natural incorporation of low-frequency radiation as e.g. catalyst for chemical change.

This work collides with the cultural burden received from social dominant ideologies epitomized by the famous Bohr-Einstein discussions;[7] this is naturally transported to ways and means in vogue helping understand the world as noted by Laughlin.[8] The representation power put over mathematical elements appearing in theoretic schemes is one example. Another pervading one being wave-particle duality. Before getting a grip on the problems encountered in this trek, one has to slowly get rid of such representations.[1-3,6,9,10]

Thus, with the representational stance discarded, the present approach permits overcoming most weird situations found so far in quantum mechanical literature; the idea is that what we call a classical world, real world so to speak, if anything, is no more than linguistic constructs; the world has always been a quantum world without being grasped as such. Quantum-classical transitions are a way of speaking to help distinguish for example emergent phenomena. [8] In simple terms, one takes the tenet proposed by Planck as a grounding principle, namely: *energy exchanged by electromagnetic radiation and matter is quantized*.

In Section 2 grounding elements concerning the present photonic approach are laid down. Section 3 examines contemporary quantum experiments from the photonic viewpoint; the aim being to familiarize the reader with the way emerging from the present approach. This section incorporates an attempt to describe atto- and femto-second chemical situations. In Section 4 a general discussion closes this work that includes a short analysis of a double-slit experiment giving an occasion to emphasize in what sense the present approach differs from views found in standard literature.

## 2. Matter-photon quantum framework

Planck (1900) studying spectra of black body radiation introduced a key idea to be found at the origin of Quantum Physics: energy exchanges in finite numbers of quanta between electromagnetic (EM) radiation and matter. The exchanged energy quantum, $\hbar\omega$, proportional to frequency $\omega$ with $\hbar$ featuring Planck's constant, namely, h divided by $2\pi$; this energy-quantum stands out as the elementary materiality tailored for q-EM. [11, 12] With fix frequency, exchange goes with $n_\omega$=0,1,2… energy clumps as the case may be. In abstract space though, frequency taken *as label*, can be changed in continuum manner; e.g. Cf. Sect.2.2. The base set appropriate to this task $\{|n_\omega\rangle\}$ leads to Fock space. [6,11,12]

2.1 Photonic bases sets

Photonic base sets obtain once electro-nuclear (EN) ones combine to Fock-space base states; they come up in two generic guises:

1) Direct product base standing for bi-partite states, e.g.: $|j\ k(j)\rangle\otimes|n_\omega\rangle$;
2) Entangled *photon-EN* basis e.g.: $|j\ k(j);n_\omega\rangle$. Here no photon energy is heralded as such for exchange; the symbol conveys non-separable matter/q-EM information.

Quantum numbers for EN base states form non-separable products of an electronic j and subsidiary nuclear-fluctuation-related quantum numbers k(j), this latter serves identifying EN-excitations of varied kinds; notation $|jk(j)\rangle$ is the simplest reminding this basic information.[1-4]

Thus, in Fock space the label $n_\omega$ identifies information on number of quantized EM energy at frequency ($\omega$) that *can be exchanged* with quantized states sustained by elementary constituent matter. This statement defines a photon concept, certainly, not as a particle; the base state symbol becomes $|n_\omega\rangle$ in abstract Fock space. The EN element contributes with labels of its own to get the symbol: $|j\ k(j);\ n_\omega\rangle$. The colored-vacuum Fock base is signaled as $|n_\omega=0\rangle \rightarrow |0_\omega\rangle$; this indicates that there is no energy quantum *available* for exchange at frequency $\omega$; yet, the corresponding *energy level* is: (1/2) $\hbar\omega$. Note that an energy level is not the energy associated to the relevant materiality; *only differences among them can be mapped to energy available at laboratory level* (Fence).

Each electronuclear base state appears in a fourfold manner constituting a basic manifold. A *resonant* electronic excitation, $E_j$ -$E_{j'}$=$\hbar\omega$, relates energy levels differing in electronic label: e.g., $|jk(j)=0\rangle\otimes|1_\omega\rangle$ and $|j'k(j')=0\rangle\otimes|0_\omega\rangle$; the energy gap relates two energy levels to one frequency value.

The elements of the basis set that are given below when arranged as generic possibilities, e.g. that for a 1-photon process mediated by an EN-system looks like:

$$(\dots |j\ k(j)\rangle \otimes |1_\omega\rangle \quad |j\ k(j);1_\omega\rangle \dots |j'\ k(j')\rangle \otimes |0_\omega\rangle \quad |j'\ k(j');0_\omega\rangle \dots ) \qquad (1)$$

The ellipsis indicates possibility to insert explicitly as many photonic bases as the problem requires. The q-state vectors are presented as amplitude sets:

$$(\dots C_{jk(j)\otimes1\omega} \quad C_{jk(j)\ 1\omega} \dots C_{j\ k(j\ )\otimes0\omega} \quad C_{j\ k(j\ )\ 0\omega} \dots )^T \qquad (2)$$

The sub-indexes refer to photonic base set elements; supplementary information can be added via amplitudes, phases and extra quantum numbers. Care is required because quantum states refer to possibilities available; they do not represent the elementary materiality, this latter can sustain infinite numbers of q-states.

Thus, the scheme includes a fix element, namely the base set vector of type (1); and a variable element corresponding to the row vector (2) that actually characterizes the q-state.

For the specific case signaled in (1), $|j'\ k(j')\rangle \otimes |0_\omega\rangle$ and $|j'\ k(j');0_\omega\rangle$ the amplitudes $C_{j\ k(j\ )\otimes0\omega}$ and $C_{j\ k(j\ )\ 1\omega}$ carry information that concerns a target state j', that here plays a role of an excited state related to ground (root) state wherefrom excitation acts; whether the excited base state is left aloof or entangled with the root colored vacuum state in the q-state signal two irreducible possibility sets that are always there. Only amplitudes affecting them might change as q-states do.

## 2.2 Space-time framework: Abstract elements

First relate quantum states belonging to the abstract level with those projected in configuration space resourcing to scalar products. Special relativity theory (SRT) provides the framework to introduce 4-dimensional (4D) space-time frames, including 3D-inertial (I)-frames.[6] The approach concerns quantum states and ways to modulate them, no representation implied.

The I-frames, besides letting in a time-axis (t) and sequential parametric ordering, permit introduction of configuration space vectors: $\mathbf{x} = (\mathbf{x}_1, \dots, \mathbf{x}_n)$ with the number 3n equaling the *total number* of classic degrees of freedom; $\mathbf{k} = (\mathbf{k}_1, \dots, \mathbf{k}_n)$ corresponds to reciprocal configuration space vectors. Vector elements are used to *label* kets; configuration space vectors: $\{|\mathbf{x}\rangle=|\mathbf{x}_1,\ldots,\mathbf{x}_n\rangle\}$; and reciprocal space vectors: $\{|\mathbf{k}\rangle = \{|\mathbf{k}_1,\ldots,\mathbf{k}_n\rangle\}$. These are sets of real numbers though no particle coordinates intended, just abstract mathematical spaces acting as functions' supports as shown later on.[6]

The amplitudes (scalar products) $\langle\mathbf{k}|\mathbf{x}\rangle$ and $\langle\mathbf{x}|\mathbf{k}\rangle$ play the role of base functions: $\exp(i\mathbf{x}.\mathbf{k})$ and $\exp(-i\mathbf{x}.\mathbf{k})$, respectively; Fourier transforms connect quantum states in a $\mathbf{k}$-basis ($g_x(\mathbf{k})$) to one in an $\mathbf{x}$-basis ($g_k(\mathbf{x})$ and vice-versa; these linear superpositions are usually named wave packets. Note that the I-frame fixes an origin to the configuration/reciprocal spaces so that the quantum states projected over these bases would share a uniform state of motion (if any) conveyed by I-frames; this is a map chosen between quantum and classical physics domains, the boundary is there.

The sets $\{|\mathbf{x}\rangle\}$ and $\{|\mathbf{k}\rangle\}$ are given the structure of rigged Hilbert spaces by introducing generalized functions, e.g. Dirac delta ($\delta$) and its derivatives: [6,12] $\langle\mathbf{x}|\mathbf{x}'\rangle\rightarrow\delta(\mathbf{x}-\mathbf{x}')$ and $\langle\mathbf{k}|\mathbf{k}'\rangle\rightarrow\delta(\mathbf{k}-\mathbf{k}')$, etc.

For the conjugate magnitudes concerning dimensions of time and frequency that, when *used as labels*, leads to base sets $|t\rangle$ and $|\omega\rangle$; thus, $\exp(i\omega t)$ and $\exp(-i\omega t)$ correspond to $\langle\omega|t\rangle\leftrightarrow\exp(i\omega t)$ and $\langle t|\omega\rangle\leftrightarrow\exp(-i\omega t)$; these scalar products play the role of basis functions for the time-frequency relationships via Fourier transforms, thereby letting relate $\omega$-space to t-space sustaining quantum states. For laboratory situations particularly tailored frequency pulses can be introduced via Fourier transforms. This is another theoretic place where laboratory space information links to the abstract framework leading to a physical photonic scheme.

Most important: Energy levels do not represent energy of the materiality; they are very useful mathematical tools (usually reflecting boundary conditions). Elimination of the representational model results in this more general view. The actual numeric value is defined to within a constant, e.g., for an I-frame in uniform motion, kinetic energy with respect to a second I-frame locating a receptor (probing) device that is to be used to detect the emission; the relative kinetic energy would shift the whole spectra (z-shift in astronomical spectroscopy [13]). The information carriers are quantum states: e.g. a photon state or a quantum state sustained by matter aggregates (antenna states for instance); quantum grounds are examined in the following section.

$\langle\mathbf{x}|\Psi,t\rangle \leftrightarrow\langle\mathbf{x}_1,\ldots,\mathbf{x}_n|\Psi,t\rangle$ stands for abstract quantum state projection on configuration space; it corresponds to a scalar product, a complex number or

complex function $\Psi(\mathbf{x},t)$ over the field of real numbers $\mathbf{x}$ and t; these fields are the functions' support. These functions *are not* the mathematical entities assumed at dawn of quantum theory by Schrödinger; what is not retained here is the representational character that was introduced via particle position space. Certainly, materiality physically sustains these quantum states though the nature of this link lies beyond experimental probe (so far); we only probe responses modulated by q-states as sketched in the preceding section.

## 2.3 Multi-partite Systems: General Basis Sets

To introduce photonic chemistry, e.g. attosecond chemistry, there is need for base states including isolated partite elements together with 1-partite states thereby simulating the possibilities for reactants and products probing. The total number of elementary material constituents is invariant.

Spectral responses permit distinguishing different types of q-states and corresponding partite terms. The quantum numbers used as labels are summarized as: $k'_1 \ldots k'_m$. The ordering of degrees of freedom is kept fix, at the end they are just numbers conveying specific information. The form of generic basis sets are given as:

i)     $\langle \mathbf{x}_1, \ldots, \mathbf{x}_m | \phi_{k\ 1 \ldots k\ m} \rangle$        (2.3.a)

ii)    $\langle \mathbf{x}_1, \ldots, \mathbf{x}_{m-1} | \phi_{k\ 1 \ldots k\ m-1} \rangle \otimes \langle \mathbf{x}_m | \phi_{k\ m} \rangle$        (2.3.b)

iii)   $\langle \mathbf{x}_1 | \phi_{k\ 1} \rangle \otimes \langle \mathbf{x}_2, \ldots, \mathbf{x}_{m-1}, \mathbf{x}_m | \phi_{k\ 2 \ldots k\ m-1\ k\ m} \rangle$        (2.3b')

iv)   $\langle \mathbf{x}_1, \ldots, \mathbf{x}_{m-2} | \phi_{k\ 1 \ldots k\ m-2} \rangle \otimes \langle \mathbf{x}_{m-1} | \phi_{k\ m-1} \rangle \otimes \langle \mathbf{x}_m | \phi_{k\ m} \rangle$        (2.3c)

A limit to these partitioning is:

x)     $\langle \mathbf{x}_1 | \phi_{k\ 1} \rangle \otimes \langle \mathbf{x}_2 | \phi_{k\ 2} \rangle \otimes \ldots \otimes \langle \mathbf{x}_{m-1} | \phi_{k\ m-1} \rangle \otimes \langle \mathbf{x}_m | \phi_{k\ m} \rangle$        (2.3x)

Even if the same number of elementary constituents sustains all q-states they are differentiated by q-number groupings and responses. To analyze them properly there is need for basis sets types identifying multi-partite state elements; these partite-states taken separately can map to fully autonomous spectral sources, e.g. $\langle \mathbf{x}_1, \ldots, \mathbf{x}_{m-2} | \phi_{k\ 1 \ldots k\ m-2} \rangle$; yet, they are not necessarily to be seen as objects in a classical sense specially when incorporated in different partitioning spaces.

As already noted, a first connection to classical world is made by associating an independent I-frame that permits introduction of configuration space as collections of real numbers, its dimension is controlled by the number of classical degrees of freedom.

Under these circumstances, there is no reasonable way to introduce a best possible classical approximation; the framework does not admit a representation mode. The quantum state drives all possible responses that include responses to coherent states; e.g. Bose-Einstein condensates.[14]

For practical reasons, one can differentiate two families:

1) All partition bases are referred to one and same I-frame, only quantum numbers are involved in identifications (labeling) including those associated to the I-frame (global rotations and translation invariances);

2) Different partitions are assigned to different I-frames albeit for some cases conserving a master I-frame (box) reference. Distance and relative orientation among I-frames can naturally be introduced according to circumstances.

Usefulness of these base sets can be sensed from the case examined below.

## 2.4 Base set for one-photon activation

The photonic multi-partite base set incorporates elements of a photon field in the same way indicated above though it becomes more specialized to the situations to be scrutinized. A base set rigged with chromophore states (label $k'_{chr}$) takes on the form:

$$|2.4a1> = <x_1,\ldots,x_{chr},\ldots x_m|\phi_{k\ 1\ldots k\ chr\ldots k\ m}>\otimes|1_\omega> \qquad (a1)$$

$$|2.4a2> = <x_1,\ldots,x_{chr},\ldots x_m|\phi_{k\ 1\ldots k\ chr\ldots k\ m};1_\omega> \qquad (a2)$$

$$|2.4a3> = <x_1,\ldots,x_{chr},\ldots x_m|\phi_{k\ 1\ldots k\ chr^*\ldots k\ m};0_\omega> \qquad (a3)$$

$$|2.4b> = <x_1,\ldots,x_{m-1}|\phi_{k\ 1\ldots k\ chr\ldots\ldots k\ m-1}>\otimes<x_m|\phi_{k\ m}> \qquad (b)$$

$$|2.4b'> = <x_1|\phi_{k\ 1}>\otimes<x_2,\ldots,x_{m-1},x_m\ |\phi_{k\ 2\ldots k\ chr\ldots\ldots k\ m-1\ k\ m}> \qquad (b')$$

$$|2.4c> = <x_1,\ldots,x_{m-2}|\phi_{k\ 1\ldots k\ chr\ldots\ldots k\ m-2}>\otimes<x_{m-1}|\phi_{k\ m-1}>\otimes<x_m|\phi_{k\ m}> \qquad (c)$$

Notation is self-evident. A coherent state would inform of quantum "distributed" over these three base elements:

$$(\ldots C_{k\ 1\ldots k\ chr\ldots k\ m\otimes1\omega}(t)\ldots\ C_{k\ 1\ldots k\ chr\ldots k\ m\ 1\omega}(t)\ldots C_{k\ 1\ldots k\ chr^*\ldots k\ m\ 0\omega}(t)\ldots$$
$$C_{k\ 1\ldots k\ chr\ldots k\ m-1\otimes k\ m}(t)\ \ C_{k\ 1\otimes k\ 2\ldots k\ chr\ldots k\ m}(t)\ldots C_{k\ 1\ldots k\ chr\ldots\otimes k\ m-1\otimes k\ m}(t)\ldots) \qquad (3)$$

The basis set order can be extracted from the sub indexes affecting the amplitudes.

Thus, only if the q-state looks like:

$$(..1_{k\ 1...\ k\ chr\ ...k\ m\otimes1\omega}\cdots0_{k\ 1...\ k\ chr^*\ ...k\ m\ 0\omega}\cdots0_{...\otimes k\ m\text{-}1\otimes k\ m}).$$

The information received is that an energy quantum might be exchanged with another q-system; otherwise, there is a lifetime associated to particular coherent q-states related to. Phase factor $\exp(\pm\mathbf{k}\bullet\mathbf{x})$ permits qualifying whether incoming or outgoing situations are intended.

In the business of basis set construction there is no interactions among partitioned sets; no representation is attempted. First, one should construct all possibilities open to the system; thereafter interactions can be incorporated; later on operators enter stage, standing either as internal operators or external sources operators or mixed).

We move on considering quantum processes related to experimental setups with the help of the above multipartite base set whenever dissociation type processes are involved. Otherwise, zero amplitude over dissociation base states permits eliminating such quantum degrees of freedom.

In the following sections we distinguish two classes of processes: physical and chemical.

## 3. Photon assisted quantum processes

The intention assigned to this section is to illustrate not only energy transfer but develop a quantum physical view of the combined photon-matter systems as they show non-separable aspects; these are essential to understand functionalities when photon states are plugged in to materially sustained states securing quantum behavior.

### 3.1. Physical processes: Reading photonic states

At resonance the photonic energy levels explicitly shown in (2) as sub indexes are degenerate. We assume ordering of EN energy levels by means of the electronic quantum number. For this case there are no jumps among the resonant multiplicity states for the simple reason that a photon-field is always incorporated into a basic fourfold bases type (1); it cannot act semi-classically also.

Thus, the semi-classic scheme depicts energy levels separated by an energy gap while a wavy arrow signals the transition; the particle picture is then apparent. Here, as just mentioned the four photonic labels disclose the same energy level value; *coupling them requires external factors* allowing for control and amplitude modulation by external fields.

Off-resonance case: e.g. energy gap $E_j$ -$E_j$ = $\hbar\omega$' > $\hbar\omega$, the probe photon-energy is smaller than the assigned gap $\hbar\omega$'. The base states $|j'k(j')>\otimes|n_\omega>$ and $|j'k(j');n_\omega>$ single out a particular EN energy level with different energy label. When $n_\omega=1$, the entangled base state shows the energy level $E_j + \hbar\omega$ lying inside $E_j$ -$E_j$ gap (by construction $\omega'>\omega$); a larger number $n_\omega$ obtains for sufficiently low frequency quantum, e.g. microwave radiation. The photonic q-state would evolve in this subspace until getting at a re-emission channel in a direction that may differ from the incident incoming photon state that originated the situation; thus, there is an in-built flexibility here .

To continue the reading exercise consider a simple example with a lambda ($\Lambda$) setup constructed around two energy levels say $E_{j1}$ and $E_{j2}$ located above level $E_{jo}$: $E_{j1}$-$E_{jo}$= $\hbar(\omega_{10})$ and $E_{j1}$-$E_{j2}$= $\hbar(\omega_{12})$; the label $j'_1$ occupies a vertex, i.e. a common target base state: $\omega_{10}>\omega_{12}$ though a subtle difference obtains via vacuum base states. If transition integral $T_{joj2}$ is zero, no direct transition is expected between "bottom" levels $j_o$ and upper $j'_2$; the $E_{j2}$–level is a sort of dark state seen from ground state $| E_{jo} >$. This system will illustrate a two-photon interaction situation.

Observe that both $|j_0, k(j_0);1_{\omega 2}>$ and $|j_2', k(j_2');1_{\omega}>$ are root states for the combined base element $| j_1', k(j_1'); 0_{\omega 10}, 0_{\omega 12}>$ their energy levels might be degenerate leading to possible coherent q-states, e.g.:

$$(\ldots C_{j0k(j0)\otimes 1\omega 20}\ \ C_{j0k(j0)\ 1\omega 20}\ \ C_{j1\ k(j1)\ \otimes 1\omega 12}\ \ C_{j1\ k(j1\ )\ 1\omega 12}\ \ C_{j2\ k(j2\ )\ \otimes 0\omega 20\otimes 0\omega 12}\ldots)^T \quad (4)$$

Observe the off-resonance states involving basis $|1_{(\omega 01-\omega 012)}>$ and ground state, namely, $|j_0, k(j_0)> \otimes |1_{(\omega 01-\omega 012)}>$ can be explicitly shown. These are possibilities to be reckoned within laboratory arrangements though not incorporated in this example since radiation $\omega_{20}$ cannot be spontaneously emitted ($T_{joj2}$=0) as a first order process.

To see the present approach at work let start up with coherent activation using the q-state (4a) playing the role of an opening channel:

$$(\ldots 1_{j0k(j0)\otimes 1\omega 10}\ \ 0_{j0k(j0)\ 1\omega 10}\ \ 0_{j1\ k(j1\ )\otimes 0\omega 10}\ldots 0_{j1\ k(j1\ )\ 0\omega 10\otimes 0\omega 12}\cdots$$
$$0_{j2\ k(j2\ )\ 0\omega 20\otimes 0\omega 12}\ \ 0_{j2\ k(j2\ )\ 0\omega 20\ 1\omega 12}\cdots 0_{j2\ k(j2\ )\otimes 0\omega 20\otimes 1\omega 12}\ )^T \qquad (4a)$$

Take a laser beam along **k**-direction at frequency $\omega_{i0}$ to encounter substrates able to sustain coherent states at the entrance channel such as:

$$(\ldots 0_{j0k(j0)\otimes 1\omega 10} \quad C_{j0k(j0)\ 1\omega 10} \quad C_{j1\ k(j1)\otimes 0\omega 10}\ldots 0_{j1\ k(j1)\ 0\omega 10\otimes 0\omega 12}\cdots$$
$$0_{j2\ k(j2)\ 0\omega 20\otimes 0\omega 12} \quad 0_{j2\ k(j2)\ 0\omega 20\ 1\omega 12}\cdots 0_{j2\ k(j2)\otimes 0\omega 20\otimes 1\omega 12}\ )^{T} \qquad (4b)$$

With the incoming channel switch off ( $0_{j0k(j0)\otimes 1\omega 10}$ ), the q-state is fully sustained by the materiality; and this can be done in infinite guises. As there is no photon energy available the situation seen from outside the I-frame can be described as stopping the laser beam; the momentum transferred is not explicitly signaled, only information remains in the base set. State (4b) can be transient since no apparent de-activation channel is present. There is then a delay time that would depend on the material and other aleatory external situations.

Interestingly, the amplitude $0_{j1\ k(j1)\ 0\omega 10\otimes 0\omega 12}$ convey the idea that we can bring up another laser beam with frequency $\omega_{i2}$ such case indicates the possibility in advance; this is a case where a basis set could be manipulated by an experimenter at discretion. Thus, taking advantage of the finite lifetime of (4b), sets up a second laser at frequency $\omega_{i2}$ and direction perpendicular to **k**, e.g. $\mathbf{k}_{\perp+}$. The q-state (4c) embodies the information concerning both actions:

$$(\ldots 0_{j0k(j0)\otimes 1\omega 10} \quad C_{j0k(j0)\ 1\omega 10} \quad C_{j1\ k(j1)\otimes 0\omega 10}\ldots C_{j1\ k(j1)\ 0\omega 10\otimes 1\omega 12}\cdots$$
$$0_{j2\ k(j2)\ 0\omega 20\otimes 1\omega 12} \quad C_{j2\ k(j2)\ 0\omega 20\ 1\omega 12}\cdots 0_{j2\ k(j2)\otimes 0\omega 20\otimes 1\omega 12}\ )^{T} \qquad (4c)$$

Note that $C_{j1\ k(j1)\ 0\omega 10\otimes 1\omega 12}$ amplitude signals the two-photon process engaging the totality of the $\Lambda$ subspace. Changing amplitude $0_{j2\ k(j2)\ 0\omega 20\otimes 1\omega 12}$ into $1_{j2\ k(j2)\ 0\omega 20\otimes 1\omega 12}$ the system opens a tri-partite contribution to the quantum state: namely $|j_2\text{'}k(j_2\text{'})\rangle\otimes|0_{\omega 20}\rangle\otimes|1_{\omega 12}\rangle$. And because amplitudes $0_{j0k(j0)\otimes 1\omega 10}$ and $0_{j2\ k(j2)\otimes 0\omega 20\otimes 1\omega 12}$ close for the time being outgoing channels the system would show up a quantum coherent state for so long they remain nigh. Thus, a form of coherence/de-coherence is explicitly included in this framework.

Decoherence can be activated if one induces a transition to e.g. state (4d):

$$(\ldots 0_{j0k(j0)\otimes 1\omega 10} \quad 0_{j0k(j0)\ 1\omega 10} \quad 0_{j1\ k(j1)\otimes 0\omega 10}\ldots 0_{j1\ k(j1)\ 0\omega 10\otimes 0\omega 12}\cdots$$
$$C_{j2\ k(j2)\ 0\omega 20\otimes 0\omega 12} \quad 0_{j2\ k(j2)\ 0\omega 20\ 1\omega 12}\cdots C_{j2\ k(j2)\otimes 0\omega 20\otimes 1\omega 12}\ )^{T} \qquad (4d)$$

State (4d) can be root-state for spontaneous emission beginning from (4e):

$$(\ldots 0_{j0k(j0)\otimes 1\omega 10} \quad 0_{j0k(j0)\ 1\omega 10} \quad 0_{j1\ k(j1)\otimes 0\omega 10}\ldots 0_{j1\ k(j1)\ 0\omega 10\otimes 0\omega 12}\cdots$$
$$0_{j2\ k(j2)\ 0\omega 20\otimes 0\omega 12} \quad 0_{j2\ k(j2)\ 0\omega 20\ 1\omega 12}\cdots 1_{j2\ k(j2)\otimes 0\omega 20\otimes 1\omega 12}\ )^{T} \qquad (4e)$$

Observe that for so long the structure of eq.(4e) does not change, there is no free photon available, so that in a way it remains "hooked" to the entangled sub space. The event, signaling detection of state $|1\omega_{12} >$ does not belong to present Hilbert space, namely:

$$(\ldots 0_{j0k(j0)\otimes1\omega10} \qquad 0_{j0k(j0)\ 1\omega10} \qquad 0_{j1\ k(j1\ )\otimes0\omega10} \ \ldots 0_{j1\ k(j1\ )\ 0\omega10\otimes0\omega12} \ldots 0_{j2\ k(j2\ )\ 0\omega20\otimes0\omega12}$$
$$0_{j2\ k(j2\ )\ 0\omega20\ 1\omega12} \ldots 1_{j2\ k(j2\ )\otimes0\omega20\otimes0\omega12} \ )^T \ \oplus \exp(i\mathbf{k}_{12}.\mathbf{R})|\ 1\omega_{12} > \qquad (4f)$$

$\mathbf{R}$ indicates detection location. This bi-partite state is the one accessible to "observers" looking at a detector screen. Yet the "click" would only identify a q-EM energy lump and the phase actor if the location of the I-frame were known.

We speak of decoherence with symbol $\oplus$ underlying this fact when laboratory energy is involved; in this case, seen at Fence a photon beam would be produced that carry information concerning direction ($\mathbf{k}_{\perp+}$), and energy ($\omega_{12}$).

We can see that the photonic framework facilitates descriptions of rather complex situations. So far only "kinematic" elements are highlighted. The scheme must be completed once specific cases are studied.

3.2 Coherence/ decoherence/ re-coherence: *Modulating speed of light*

Let examine some reported observations of coherent optical information storage [15] from the perspective offered by the present approach.

The experimental set up includes three laser sources so that the $\Lambda$-model from section 2.1 provides a basis for discussion. Again at frequency $\omega_{20}$ a beam moving in direction $\mathbf{k}_{forward}$ so that interaction with the materiality takes place. A second laser activates at frequency $\omega_{21}$ in directions perpendicular to the first, say $\mathbf{k}_+$; result: the beam $\mathbf{k}_{forward}$ stops. A third laser can be activated after a while from direction $\mathbf{k}_-$ i.e. in the direction opposite to $\mathbf{k}_+$. What do we observe?

First, stop the energy pulse by inducing a dark transition using the second beam, $\mathbf{k}_+$. Wait, thence set up the pulse $\mathbf{k}_-$: one observes the initial pulse revival! [15]

The following quantum states schematically sum up the experimental response:

Step 1 corresponds to a laser set up at resonance frequency $\omega_{20}$ beginning quantum interactions with the materiality sustaining the base states indicated as sub-indexes; the forward direction $\mathbf{k}_{forward}$ is left implicit to alleviate the writing. Preparation:

$$1: (\ldots 1_{j0k(j0)\otimes 1\omega 20} \quad 0_{j0k(j0)\,1\omega 20} \quad 0_{j1\,k(j1)\otimes 1\omega 12} \quad 0_{j1\,k(j1)\,1\omega 12}\ldots 0_{j0k(j0)\otimes(1\omega 02-\omega 12)} \quad 0_{j0k(j0)\,1(\omega 02-\omega 12)}$$
$$0_{j2\,k(j2)\,0\omega 20\otimes 1\omega 12} \quad 0_{j2\,k(j2)\otimes 0\omega 20\otimes 0\omega 12} \quad 0_{j2\,k(j2)\,0\omega 20\,1\omega 12} \quad 0_{j2\,k(j2)\,0\omega 20\otimes 1\omega 12}\ldots)^T \qquad (5a)$$

A new q-state can be open via entanglement. In this case, photon state entanglement,

$$2: (\ldots C_{j0k(j0)\otimes 1\omega 20} \quad C_{j0k(j0)\,1\omega 20} \quad 0_{j1\,k(j1)\otimes 0\omega 02} \quad 0_{j1\,k(j1)\,0\omega 02}\ldots 0_{j0k(j0)\otimes(1\omega 02-\omega 12)} \quad 0_{j0k(j0)\,1(\omega 02-\omega 12)}$$
$$0_{j2\,k(j2)\,0\omega 20\otimes 1\omega 12} \quad 0_{j2\,k(j2)\otimes 0\omega 20\otimes 0\omega 12} \quad 0_{j2\,k(j2)\,0\omega 20\,1\omega 12} \quad 0_{j2\,k(j2)\,0\omega 20\otimes 1\omega 12}\ldots)^T \qquad (5b)$$

In this step, amplitudes develop corresponding to a coherent interaction between the incoming laser pulse $\mathbf{k}_{forward}$ and the materiality including excited state located above ground state with energy label $E_{j0k(j0)} + \hbar\omega_{20}$. An internal (photonic) coherent state reads as:

$$3: (\ldots 0_{j0k(j0)\otimes 1\omega 20} \quad C_{j0k(j0)\,1\omega 20} \quad 0_{j1\,k(j1)\otimes 0\omega 20} \quad C_{j1\,k(j1)\,0\omega 20}\ldots 0_{j0k(j0)\otimes(1\omega 02-\omega 12)} \quad 0_{j0k(j0)\,1(\omega 02-\omega 12)}$$
$$0_{j2\,k(j2)\,0\omega 20\otimes 1\omega 12} \quad 0_{j2\,k(j2)\otimes 0\omega 20\otimes 0\omega 12} \quad 0_{j2\,k(j2)\,0\omega 20\,1\omega 12} \quad 0_{j2\,k(j2)\,0\omega 20\otimes 1\omega 12}\ldots)^T \qquad (5c)$$

This coherent state necessarily shows up lifetime, as no emission channel is open; and, while it is "alive", activate a second laser direction $\mathbf{k}_+$ that is perpendicular to $\mathbf{k}_{forward}$; this beam interacts with the preceding state, namely via $C_{j1\,k(j1)\,0\omega 20}$. The frequency chosen so as to get a coherent state incorporating the level $E_{j2\,k(j2)}$. This state activated from $C_{j1\,k(j1)\,0\omega 20} \neq 0$ was open by the first laser. The laser momentum included in the coherent state may possibly propagate e.g.:

$$4: (\ldots 0_{j0k(j0)\otimes 1\omega 20} \quad 0_{j0k(j0)\,1\omega 20} \quad 0_{j1\,k(j1)\otimes 1\omega 12} \quad C_{j1\,k(j1)\,1\omega 12}\ldots 0_{j0k(j0)\otimes(1\omega 02-\omega 12)} \quad 0_{j0k(j0)\,1(\omega 02-\omega 12)}$$
$$C_{j2\,k(j2)\,0\omega 20\otimes 1\omega 12} \quad C_{j2\,k(j2)\otimes 0\omega 20\otimes 0\omega 12} \quad 0_{j2\,k(j2)\,0\omega 20\,1\omega 12} \quad 0_{j2\,k(j2)\,0\omega 20\otimes 1\omega 12}\ldots)^T \qquad (5d)$$

Note: energy levels associated to $C_{j1\,k(j1)\,1\omega 12}$ measured from the dark level $C_{j2\,k(j2)\otimes 0\omega 20\otimes 0\omega 12}$ corresponds to two photons $n_{\omega 12}=2$. Thus, once the beam $\mathbf{k}_+$ goes through it carries one photon in excess, produced by induced photon emission (not shown). The physical time evolution may result from opening of decoherence channel:

5: $(\dots 0_{j0k(j0)\otimes 1\omega 20} \quad 0_{j0k(j0)\ 1\omega 20} \quad 0_{j1\ k(j1\ )\otimes 1\omega 12} \quad 0_{j1\ k(j1\ )\ 1\omega 12} \dots 0_{j0k(j0)\otimes (1\omega 2\text{-}1\omega 12)} \quad 0_{j0k(j0)\ 1(\omega 02\text{-}\omega 12)}$
$0_{j2\ k(j2\ )0\omega 20\otimes 1\omega 12} \quad 0_{j2\ k(j2\ )\otimes 0\omega 20\otimes 0\omega 12} \quad 0_{j2\ k(j2\ )\ 0\omega 20\ 1\omega 12} \quad 1_{j2\ k(j2\ )\ 0\omega 20\otimes 1\omega 12} \dots)^T$ \hfill (5e)

The q-state (5e) once emission takes place leads to (5e'):

$(\dots 0_{j0k(j0)\otimes 1\omega 20} \quad 0_{j0k(j0)\ 1\omega 20} \quad 0_{j1\ k(j1\ )\otimes 1\omega 12} \quad 0_{j1\ k(j1\ )\ 1\omega 12} \dots 0_{j2\ k(j2\ )\ 1\omega 12} \quad 1_{j2\ k(j2\ )\otimes 0\omega 12} \dots)^T$ \hfill (5e')

A memory loss with respect to state (5e) happens that is understandable when one wants to prepare the state corresponding to (5e). Thus, if only one photon $\omega_{12}$ is available, there is no way to get back to state (5a) because frequency has to come up as $\omega_{20}$.

A second photon at frequency $\omega_{20}$ is required so that using the virtual state $(1\omega_{02}\text{-}1\omega_{12})$ is brought up to $1\omega_{02}$ and the frequency up-conversion holds. From this point onwards the process can be driven to get state (5a). To accomplish this fate, a laser beam is sent back with direction $\mathbf{k}_-$. As linear momentum is conserved, these photon-states take back the direction $\mathbf{k}_{forward}$.

The result after a finite time-delay there will be a flash in direction $\mathbf{k}_{forward}$!

Halted light pulses, similar to those described above, prepared by Liu et al. [15] permits storage of coherent optical information in atomic materials. Actual implementation is quite sophisticated in so far use of coupling and probe lasers handling is concerned. A careful analysis of this reference would help readers to sense the import of such experiments.

Such is the beauty of linear momentum conservation. So that one can tinker with the time delay imposed onto the laser pulse in the forward direction. One can describe this situation by assigning a velocity to the laser pulse much smaller than the speed of light. Actually one is playing with coherence-decoherence and re-coherence phenomena.

Thus, Coherence and de-Coherence are essential laboratory phenomena. Re-coherence is a special operation leading to a desired intermediate state. Note that by putting zero to particular amplitudes one "erases" the possible response from the given base state. One can erase a response though not a base state; these are always present thereby supporting emergence phenomena.

So far, the presentation avoids entanglement aspects.[10] Some problems are discussed below; these concern chemistry in the first place.

## 3.3 Chemical processes

i) Competitive chemical reactivity in 1-photon field

Consider a material system able to sustain spectral responses that present a particular chromophore-manifold; this one might modulate outcomes eventually related to chemical processes.

The chromophore set (chr) identifies with base states label: $\{|j_{ch}'',k(j_{ch}'')>\}$; the first electronic excited label $E_{(jch\ =1,k(jch\ =1)=0)}$ locates well above possible dissociative energy channels measured with respect to the 1-partite system. This illustration includes two generically labeled channels: $|B^1>$ and $|B^2>$. Definitions:

$$|k'_1 k'_2 k'_{ch} k'_{m-1} k'_m > \otimes |0_\omega > \longrightarrow |A^0_{k\ 1\ k\ 2\ k\ ch\ k\ m-1\ k\ m \otimes 0\omega} >$$

$$|k'_1\ k'_2 k'_{ch} k'_{m-1} > \otimes |k'_m > \quad \longrightarrow |B^1_{k\ 1\ k\ 2\ k\ ch\ k\ m-1 \otimes k\ m} >$$

$$|k'_1 > \otimes |k'_2 k'_{ch}\ k'_{m-1} k'_m > \longrightarrow |B^2_{k\ 1 \otimes k\ 2\ k\ ch\ k\ m-1,k\ m} >$$

The bipartite channels have no indication of external q-EM fields. This is done to focus discussion onto the dissociation direction as modulated by chromophore activation.

The base vector below permits discussing aspects of the phenomenology associated to some experimental models using a 1-photon state case:

$$|A^0_{k\ 1\ k\ 2\ k\ ch\ k\ m-1\ k\ m} > \otimes |1_\omega >; \text{ ingoing/outgoing possibility channels.}$$

$$|A^0_{k\ 1\ k\ 2\ k\ ch\ k\ m-1\ k\ m \otimes 1\omega} >; \text{ photon induced interaction possibilities.}$$

$$|A^0_{k\ 1\ k\ 2\ k\ ch\ k\ m-1\ k\ m\ 1\omega} >; \text{ photon entangled possibilities.}$$

The photon-entangled base set does not localize energy quanta, thus from a coherent state there cannot be emission unless time evolution, induced by external action, leads to such amplitudes opening window channels.

Quantum states expressed in these photonic bases following the eq.(1) pattern read:

$$(C^{A0}_{k\ 1;k\ ch;k\ m \otimes 1\omega}(t-t_o) \ldots C^{A0}_{k\ 1;k\ ch;k\ m;1\omega}(t-t_o) \ldots$$
$$C^{B1}_{k\ 1\ k\ 2\ k\ ch* k\ m-1 \otimes k\ m;0\omega}(t-t_o) \ldots C^{B2}_{k\ 1 \otimes k\ 2\ k\ ch* k\ m-1\ k\ m;0\omega}(t-t_o) \ldots) \quad (6)$$

Consider at $t = t_o$ the system prepared in the following window q-state:

$$(1^{A0}_{k\ 1\ k\ 2\ k\ ch\ k\ m-1\ k\ m \otimes 1\omega}(t_o) \ldots 0^{A0}_{k\ 1\ k\ 2\ k\ ch\ k\ m-1\ k\ m;1\omega}(t_o) \ldots$$
$$\ldots 0^{B1}_{k\ 1\ k\ 2\ k\ ch* k\ m-1 \otimes k\ m;0\omega}(t_o) \ldots 0^{B2}_{k\ 1 \otimes k\ 2\ k\ ch* k\ m;0\omega}(t_o)) \quad (6a)$$

Note the 1-photon energy exhausts in forming this initial state. This hint originates from this formalism. To continue the process one has to induce time evolution. In Hilbert space we need no more energy except for an interaction Hamiltonian.

In laboratory space external sources are require to produce physical effects. By hypothesis let us assume we have all what is required to continuing state evolution.

For $C^{A-0}$ the negative super index targets the infinite bunch of initial states an experimenter may prepare. Conventionally, the negative super-index attaches to momentum base states $\exp(-i\mathbf{k}\bullet\mathbf{x})$ while a positive one does it for $\exp(+i\mathbf{k}\bullet\mathbf{x})$; the origin of the material system I-frame is implied. Spontaneous emission may implicitly relate to state (6a); i.e. material system is to be seen as a source of a photon state provided free photon states are released.

The experimenter prepares the system by focusing a laser beam or simply a light beam with frequency in resonance with the chromophore first excited state. Entanglement between photon and matter base states retains EM energy to form a photonic system, e.g.,

$$(C^{A\pm0}{}_{k\,1\,k\,2k\,ch\,k\,m\text{-}1\,k\,m\otimes1\omega}(t\text{-}t_o)\dots C^{A0}{}_{k\,1\,k\,2\,k\,ch\,k\,m\text{-}1k\,m\,1\omega}(t\text{-}t_o)\dots$$
$$0^{B1}{}_{k\,1\,k\,2\,k\,ch^*\,k\,m\text{-}1\otimes k\,m\,0\omega}(t\text{-}t_o)\dots 0^{B2}{}_{k\,1\otimes k\,2\,k\,ch^*\,k\,m\text{-}1\,k\,m\,0\omega}(t\text{-}t_o)) \tag{6b}$$

Observe that as soon as coherence dissipates, it is the information put for vacuum field that is suppressed:

$$(0^{A\pm0}{}_{k\,1\,k\,2k\,ch\,k\,m\text{-}1\,k\,m\otimes1\omega}(t\text{-}t_o)\dots C^{A0}{}_{k\,1\,k\,2\,k\,ch\,k\,m\text{-}1k\,m\,1\omega}(t\text{-}t_o)\dots$$
$$C^{B1}{}_{k\,1\,k\,2\,k\,ch^*\,k\,m\text{-}1\otimes k\,m}(t\text{-}t_o)\dots C^{B2}{}_{k\,1\otimes k\,2\,k\,ch^*\,k\,m\text{-}1\,k\,m}(t\text{-}t_o)) \tag{6b'}$$

A coherent state (6b') obtains engaging the chromophore subspace only. Such state would appear characterized by a lifetime.

Although as soon as forced time evolution would lead to state (6a), possibility opens for decoherence through a 1-photon state emission as indicated above.

One speaks decoherence because the base set element is not contained in the photonic base states, it only has information of the source (i.e. I-frame). If the photon state is actually sensed it does it via an event (click) or some other irreversible process that can be elicited later on.

Summarize some generic possibilities producing or not chemical changes including laboratory situations:

1) Re-emission of a photon state in any direction leaving the 1-partite system in a ground (root) state.

2) If the incoming photon state shows a somewhat larger frequency than the 0-0 transition then low energy photon states from the chromophore can give off low frequency radiation that eventually can be used to identify changes of the chromophore itself;

3) Filter radiation so as to produce a monochromatic 0-0 frequency in case of emission.

4) Open up couplings to dissociative channels, e.g. $B^1$ and/or $B^2$.

ii) A simple quantum chemical computation

To make contact with chemically oriented schemes consider the following simple model eq.(7). These matters can be examined with the help of an effective four-electronic states model would produce a useful computation model [1] schematically shown below.

Set up a four-state model embodying a lambda sub space ($E_1$, $E_2$, $E_3$):

$A^0_{k \, 1k \, ch \, k \, m-1k \, m \, 1\omega} \rightarrow E_0$; $A^0_{k \, 1 \, k \, 2 \, k \, ch^* \, k \, m-1k \, m} \rightarrow E_1$; $E^{B1}_{k \, 1 \, k \, 2 \, k \, ch \, k \, m-1 \otimes k \, m} \rightarrow E_2$, with

$E_2 = E_{k \, 1 \, k \, 2 \, k \, ch \, k \, m-1} + E_{k \, m}$; $E^{B2}_{k \, 1 \otimes k \, 2 \, k \, ch^* k \, m-1k \, m} \rightarrow E_3 = E_{k \, 1} + E_{k \, 2 \, k \, ch \, k \, m-1 \, k \, m}$ (7)

Understood, take the lowest compatible set of indexes to set up the model; the following inequality holds under this caveat: $|E_1 - E_2| < |E_1 - E_3|$.

Examine the generic quantum state:

$$|\Phi> = C_0 \, |E_0> + C_1| \, E_1> + C_2| \, E_2 > + C_3 | \, E_3>$$  (8)

Including interaction Hamiltonian between the base set elements with low frequency fields and solving the associated secular equation for the root level found nearest to the excited state (ch*) one gets for the amplitudes the inequalities $|C_1| > |C_2| \gg |C_3|$. These results obtain by solving well-known secular equations. [1]

One can safely conclude that response from channel 2 will show up before the one from channel 3 under the excitation conditions implied.

Now, if activation were started up from ground state, then gaps are inverted the conclusion is obviously the opposite. The response from channel 3 will precede the one from channel 2. Namely, normal chemistry would show up.

Under steady illumination condition the partitioning favors channel 2 if the chromophore responds to the photon field channeling the energy to the material system. The photochemistry can hence be apprehended in a nutshell. Note that a 1-photon process is not chemically effective. One would stop at the coherent state at best if driving fields were switch off.

This overview helps sensing the power of the photonic scheme. But there are other issues in the photonic scheme, some of which we discuss below.

3.4. Entanglement: Chemical entanglements

Quantum entanglement concerns quantum states; entangled states sustained by elementary material systems but this latter is not driven by. There is no place for a classical picture.[3]

 Spin degrees of freedom are included in what follows. Bearing in mind bases of spinorial nature, 2-dimensional base vectors must be consider for spin $S=\frac{1}{2}$: ($|\alpha>$    $|\beta>$). A spin q-state reads as: ( $C_\alpha$    $C_\beta$)$^T$. These bases are combined with space dependent partite states, e.g.,

$$\{<x_1,\dots,x_{ch},\dots,x_{m-1}|\phi_{k\ 1\dots k\ ch\dots k\ m-1}> (|\alpha>_{m-1}\ |\beta>_{m-1})\}\otimes(|\alpha>_m\ |\beta>_m) <x_m|\phi_{k\ m}>$$
$$(9a)$$

A second possibility concerns label permutation that reads:

$$\{<x_1,\dots,x_{ch},\dots,x_{m-1}|\phi_{k\ 1\dots k\ ch\dots k\ m}> (|\alpha>_m\ |\beta>_m)\}\otimes (|\alpha>_{m-1}\ |\beta>_{m-1}) <x_m|\phi_{k\ m-1}>$$
$$(9b)$$

The spinorial character reflects now on the partite base states. Sub indexes of the first spinor (quantum numbers) associates to the large partite base set, the second spinor to the 1-partite contribution. This introduces a "locality" idea.
In a coherent situation both are referred to the same I-frame; if each one associates to a particular I-frame the system would be mapped to possibilities that can be found for colliding pairs.

The possibilities signaled above indicate permutations of equivalent q-labels. Spin tensor product leads to spin singlet component:

$$F(S=0) = 1/\sqrt{2} \ (|\alpha>_{m-1}|\beta>_m - |\beta>_{m-1}|\alpha>_m) \times$$
$$1/\sqrt{2} \ ( \ |\phi_{k\ 1\dots k\ ch\dots k\ m-1}> |\phi_{k\ m}> + |\phi_{k\ 1\dots k\ ch\dots k\ m}> |\phi_{k\ m-1}>) \quad (10)$$

Spin triplet multiplet components ($M_S=0$):

$F^+(S=1) = 1/\sqrt{2} \, (|\alpha>_{m-1}|\beta>_m + |\beta>_{m-1}|\alpha>_m) \times$
$$1/\sqrt{2} \, ( \, |\phi_{k \; 1\ldots k \; ch\ldots k \; m-l}> |\phi_{k \; m}> - |\phi_{k \; 1\ldots k \; ch\ldots k \; m}> |\phi_{k \; m-1}>) \qquad (11a)$$
For $M_S=1$:
$F^{++} = |\alpha>_{m-1}|\alpha>_m \, 1/\sqrt{2}( \, |\phi_{k \; 1\ldots k \; ch\ldots k \; m-l}> |\phi_{k \; m}> - |\phi_{k \; 1\ldots k \; ch\ldots k \; m}> |\phi_{k \; m-1}>) \; (11b)$
For $M_S= -1$:
$F^{--} = |\beta>_m |\beta>_{m-1} 1/\sqrt{2}( \, |\phi_{k \; 1\ldots k \; ch\ldots k \; m-l}> |\phi_{k \; m}> - |\phi_{k \; 1\ldots k \; ch\ldots k \; m}> |\phi_{k \; m-1}>) \qquad (11c)$

Only amplitudes associated with quantum numbers are changing. If one applies permutations to configuration space coordinates the illusion of particle entanglement may occur. Yet the way we handle spin implies a kind of localization that is a characteristic of chemical systems.

$F^-$ and $F^+$ in (10)&(11a) present spin and space entanglement; while $F^{++}$ and $F^{--}$ in (11b)&(11c) show space entanglement only with a nodal plane between permuted quantum numbers. In chemical language these correspond to "radical" states (bi-radical in particular).

 Note that for $F^-$ the permutations of space labels, the space component do not change sign; it is symmetric, and in chemistry this quality is usually related to bonding concepts; while it is the spin function that changes sign ensuring fermionic character.

Spin labels permutation leave invariant $F^+$, $F^{++}$ and $F^{--}$. These latter functions complete the spin triplet subspace when cast into molecular physics language. The space function, however, does change sign under label permutation. In chemistry such space function corresponds to anti-bonding concepts. Obvious from this perspective, the triplet state would play an essential role in mediating changes of states.[4]

The introduction of spin degrees of freedom, for this generic case, opens the possibility to entangled bases in a quantum physics manner.

Interestingly, a general property relating 1- to multi-partite states corresponds to emergence of nodal planes between related partite sections.
The role of a spin triplet state in preparing conditions to changing amplitudes from a 1-partite to a bipartite state is hence a quantum physical requirement not a mechanical stretching one. The electronic quantum numbers including spin control possible fragmentation; this is a generic property.

Furthermore, entanglement seems to be a requisite to *move amplitudes* from a 1-partite state into a bi-partite state. There is no classical mechanical bond breaking or knitting along nuclear vibration coordinates; just connections between quantum possibilities driven by electronic amplitude variations.

For I-frames the anti-bonding case introduces the concept of nodal plane separating the configuration space coordinates and in the case where a bi-partite system is involved the nodal plane would separate in real space the systems associated to each fragment; similarly for multi I-frame systems. In this context, distance concept enters in all correctness. Yet, there is a caveat involving decoherence. Without this latter phenomenon and recoherence chemical change could not be seen as a quantum physical phenomenon.

The couple decoherence/recoherence plays central roles in apprehending the quantum physical nature of bond knitting/breaking in chemistry. This is a rewarding result.

### 3.5 Atto-chemistry

What type of chemistry one would expect once a standard molecular system is submitted to attosecond laser pulses? The question has been partially examined by us [2-4] as well as with the help of semi-classic quantum chemical procedures by others.[16,17]
Attosecond pulses show maximum amplitude centered at say energy level $\omega_o$ corresponding to a very high frequency; energy available that may be well above ionization and/or dissociation limits referenced to a 1-partite ground (root) state or some other state of interest. The pulse width $\delta\omega$ being very much smaller than $\hbar\omega_o$ appears to be quantized, e.g. harmonics levels to be found above and below the energy level value $\hbar\omega_o$.

Interaction of an attosecond pulse and ground state sustained by a given elementary materiality can be registered as coherent state:

$$(\dots C_{j0k(j0)\otimes 1\omega}\, C_{j0k(j0)\,1\omega}\dots \underline{C_{j0\ k(j0)\otimes 1\omega - n\omega}}\ \underline{C_{j0\ k(j0)\otimes 1\omega - n\omega + 1\omega}}\dots \underline{C_{j0\ k(j0)\otimes 1\omega \pm 0\omega}}\,\cdots$$
$$\underline{C_{j0\ k(j0)\otimes 1\omega+1\omega}}\dots C_{j1\ k(j1)\otimes 0\omega +n\omega}\dots 0_x\dots)^T \qquad (12)$$

The set $0_x$ stands for a host of EN energy levels found degenerate to those in the energy band carried by the attosecond pulse; among excited EN states there are all those that would open channels to varied products and intermediate q-states. Q-state (12) plays the role of initial state.

Thus a pulse is initially imprinted onto the EN q-state (underscore in (12)); seen from the Fence energy would be shared with the materiality sustaining the q-state. This initial state would evolve in Hilbert space, meaning with this statement that all decoherence channels are switched off.

Note that there are no free photons at disposal and consequently no residual EM field. This latter in the photonic framework cannot be active to drive evolution in time and at the same time pay the energy bill; this latter only makes sense in the semi-classic scheme.

To "break" this initial state stalemate and change the amplitudes there is need of external low frequency radiation that would induce e.g. time evolution thereby providing mechanisms for grasping emergence poly-partite states.

So long the system is trap in coherent states there is no free-electron and/or ions wandering in space that could be trap back by a putative ionized core.

One confronts a situation unedited in standard QM. The interaction Hamiltonian must be able to move real energy such as the expended in producing EM fields. One way to induce time evolution at the Fence goes through shinning low frequency (e.g. microwave) radiation. This supplement acts as a catalyst. At this point, and place (Fence) low-frequency fields might be included in a semi-classic manner.

Mechanisms of decoherence can now be inserted that would set up free I-frames partite q-states including free radicals, proton and/or electron states.

The opportunity to implement simulations frameworks is now at hand. An example is given in ref.[1] As the framework does not represent materiality, inclusion of electric and magnetic fields in a classic guise must be done via semi-classic formulation. Thus, time evolution at the Fence differs from the one found in abstract space.

In the present context, the roles of infrared, microwave or even much lower frequency radiation are important to propagate a chemical change signaled by non-zero amplitudes over electronic quantum number that can eventually be relaxed via cascades of k(j) quantum numbers to chemical species differing from the initial one.[1]

The quantum physical nature of not only chemical but also biologic processes becomes apparent.

4. Discussion

A photonic framework developed in full q-physical terms as implemented here lets no obvious place to incorporate methods and ideas from advanced quantum chemical schemes such as potential energy hyper-surfaces. It is then complementary to all what is practiced in theoretical chemistry. At laboratory level actual electromagnetic systems enter semi-classically to drive q-change thereby accomplishing the chemical process; the choice of radiation pends on the chemist requirement. It is this dynamical side that belongs to both domains, classical and quantum that is not reducible so as a "theory of everything" is not possible.

The approach recasts quantum physics in terms of states that are sustained (not represented) by basic material elements (for chemistry: electrons and nuclei). At boundaries the approach must be kept open to incorporate semi-classic EM fields acting as "motor fuel" to help drive real chemical setups. Non-unitary time evolution at quantum physical space boundaries is a necessary characteristic. Decoherence, recoherence and coherence are quantum phenomena that would break the simplicity of closed systems imposing limitations to unitary time evolution. This latter is used to explore possibilities though not to represent processes. Its lack relieves the abstract scheme to represent actual processes thereby opening an opportunity window to implement physical approaches at the Fence; we must consider energy-entropy costs in the long run. Possibilities become restricted and at the boundaries within detectors transformed into events; counting these events permits introducing probability measures.

Entanglement is a central issue unconcealed by quantum physics. The introduction of a different manner to probe the nature of quantum states permits relating abstract to laboratory levels. Quantum mechanical weirdness has been expelled as much as it is possible.[2,3] On the one hand, by replacing the representational mode: one never follows the "paths" of particles, so that there might be quantum states sustained by material elements only.

On the other hand, language reflecting classical tenets has been cleaned up to the best of our abilities: wave/particle *duality* does not make sense. Quantum phenomenon is what there is; thus one better has to develop a language adequate to circumstances to get descriptions either in prose or poetry. Oxymoronic expressions must be expelled from the present way we use to describe quantum phenomenology. Realistic pictures also.

A connection between "clicking" and quantum entangled states can be advanced. Besides the "mechanical" aspect (interaction with detectors) emphasized by not only the present approach, a concomitant aspect is the expression of the whole quantum state that is meaningful. [2-5] An analysis from the laboratory perspective exalts the collision aspect of the interaction. The present approach exalts the global quantum nature that such events unravel. Both schemes appear to be incomplete. If there is complementarity, it must be at the level of apprehension of quantum events. Even if "clicking" sounds classical in last resort it is just a quantum event. What appears to be classical might be the amplification device used at laboratory level.

Present day physics equate an event having been caused with predictability of such an event. Both levels are not commensurate: the former is ontological (principle of causality); the latter (predictability) is epistemological. Even today there are people discussing these issues without realizing their intrinsic *incommensurateness*.

Grounding issues around quantum physics as well as physical chemistry have been important targets for this paper. The classical principle of locality states that an *object* is influenced directly only by its immediate surroundings and not by remotely located objects. In this sense, quantum physics would appear to be manifestly non-local; but there is a problem here. Actually, in quantum physics there are no objects: only quantum states sustained by basic material constituents. Deceitfully, use of classical concepts distorts the formulation concerning quantum ones. Local/non-local concepts suffer from this ambiguity and in the long run must be modified or even removed. One better employ separability/non-separability concepts. Entanglement would conceivably exemplify non-separability rather than non-locality; a separable element defines a partite system. The elementary materiality does support quantum states in a fully non-classical way to the extent that there is no classic bonds instantiating sustainment. Bohr's influence is not helping to clear up this issue; [7] one can consider with Heidegger [18] that "*everything decisive is despite the ordinary, for the ordinary and usual recognizes and wants only its own kind*" (Cf. p.35). In this respect, a brief comment on the double-slit experiment seems in place. The setup introduces definite semi classic features: double slit location in laboratory space is one. The initial quantum state $|\Phi>$ *interacts* with both scattering center and new possibilities develop; e.g.: $\hat{V}_1 |\Phi>$ and $\hat{V}_2 |\Phi>$, that except for location of the I-frames both states are identical. Beyond the plane including the slits and normal to it the quantum state reads: $C_1 \hat{V}_1 |\Phi> + C_2 \hat{V}_2 |\Phi>$; it describes all possibilities the quantum system may show up after slit-interactions. In between the slits,

$C_1 = C_2 = 1/\sqrt{2}$ the function (projected q-state) presents a maximum. And the "picture" formed one-by-one by the events corresponds to an interference pattern. Thus the initial state plus the two components *reflecting the interaction with the slits* forms the quantum state. There is no "interpretation" in terms of "which path" took a particle that would be "represented by |Φ>. Any manipulation made at the slits to determine passage of materiality would directly affect the amplitudes thereby producing $C_1 \neq C_2$ so that interference pattern fades away.[5] Because the materiality sustains the q-state, the necessary and sufficient condition is that the materiality attains the detecting plane; otherwise, there is nothing. The way it does is not one of the business of quantum physics simply because the question addresses an ordinary classical physics issue. This is the usual situation and *the ordinary and usual recognizes and wants only its own kind*.

The discussions presented thus far addresses basic problems in quantum physics and quantum mechanics in particular. In the opinion of Steven Weinberg there is no entirely satisfactory interpretation of quantum mechanics;[19] Hobson's [20] experiments blur even more the discussion with the analysis of two-photon interferometry; the usual basic elements remain here: Particle model and representational mode are still there. Our premises on the other hand: "at the laboratory level material basic elements (electrons and/or nuclei) *sustain* quantum states yet these quantum states do not represent objects; and objects (particles) do not occupy base states"; these ideas do replace the dominating paradigm. Though it does not offer a "picture", yet. It suffices to realize that the linear superposition is sustained by the materiality as suggested by the double slit just discussed; but there are no objects to occupy base states. It is to the reader to think through this problem and make comparisons.[6] Though it is an open problem it seems that the hitch is with the meaning of interpretation.

The present viewpoint favors the following stance: there is no need for interpretations in classical physics terms.

Yet, event sequences as such, which are recorded at laboratory premises, can be submitted to probabilistic analyses; they share an "objectiveness" character related to decoherence, namely quantum decoherence. In this case the weight is put on the recorded material: the spot (click). It is then reasonable that Bayesian analyses be useful in this context as the enquire direction is from outside (laboratory) to the inside (quantum). However, the supporting role of materiality with respect to quantum states, that is fundamental for the present view, fades away in a probabilistic (counting) context. And, consequently the weirdness of quantum mechanics would again be apparent to the extent we insist in talking classical physics language to describe quantum experimental setups and outcomes.

To bypass such contradictory way of speaking we ought to accept that Classical physics and Quantum physics languages are, strictly speaking, irreconcilable. Yet they can still be used in harmonizing manners with the classical ones playing a subsidiary role, the introduction of inertial frames is one example.[2-6] Thus, simulation procedures balancing abstract with semi-classic quantum schemes may shed new light on a number of physical-chemical-biological processes when the present context helps extracting the relevant information. Thus, a recorded interference pattern is just a q-state taken as a whole, namely, cumulating a minimal amount of events for us to see "the picture". Tonomura experiments [21] provide beautiful illustrations of what is at stake concerning q-state as a whole yet energy exchange with a sensing screen is local. The latter is not the characteristics for this situation: any one of such spots makes part of the whole q-state yet elicit real energy exchange.[5,6]

Coherence is a key concept that becomes simply formulated in the photonic framework. The size of the materiality sustaining quantum states is not a restrictive element. The classic view is not apodictic though it might help us that are trying to get on the way to apprehending quantum phenomena. Clearly, it cannot help understand the quantum phenomenon itself; so far the wave-particle duality has blurred the key issue only. [5] Brumer and Shapiro [22] reported early studies of one photon mode selective control of reactions by rapid or shaped laser pulses. A specific competitive chemical reactivity in 1-photon field examined above elicits the role of a chromophore in guiding (so to speak) the selection of products; the reaction control being mediated by a "third body", namely, the chromophore spectrum.

The EN q-state prevents localizations of nuclear positions; the concept of structure, be it of a transition state structure loses preeminence. Yet it may help "translate" the nature of resonances discussed above taken as electronic quantum numbers or labels; it may help giving names orienting the discussions.[2-4]

The catalytic role of microwave radiation is understandable in the photonic view following lines introduced here; low frequency excitations find a natural place in this approach as both energy carrier and coupling element. As a matter of fact, another simulation research domain opens; this can help unlock computation black boxes to practitioners.

The difficulties found in the scheme when abstract and Fence states come together were examined. The result: these end processes require of energy conservation. At the same time they provide supplementary driving sources to ensure time evolution. This is a price one pays when open systems enter the discussion. Actually, it is their strength.

Acknowledgements

The author thanks all coworkers that have helped develop these ideas. In particular Prof. R. Contreras (Chile University), G. Arteca (Laurentian, Sudbury, Canada), JM. Aulló (Valencia University) and E. Brändas (Uppsala University).